\documentclass[%
 reprint,
 amsmath,amssymb,
 aps,
 prl,
 floats,
]{revtex4-1}

\usepackage{subfigure}
\usepackage{graphicx}
\graphicspath{{./}{./images/}{./images_all/}{./images_all/thermal/}}
\DeclareGraphicsExtensions{.png}

\usepackage{dcolumn}
\usepackage{bm}

\begin{document}

\preprint{APS/123-QED}

\title{Coherent spin transport and suppression of spin relaxation in InSb nanowires with single subband occupancy
at room temperature}

\author{Saumil Bandyopadhyay$^{1,3}$$\thanks{Present address: Massachusetts Institute of 
Technology, Cambridge, MA 02139.}$, Md. Iftekhar Hossain$^2$, Hasnain Ahmad$^2$, Jayasimha Atulasimha$^3$
and Supriyo Bandyopadhyay$^2$}
\email{sbandy@vcu.edu}
\affiliation{$^1$Maggie L. Walker Governor's School for Government and International Studies, 
Richmond, VA 23220, USA \\
$^2$Department of Electrical and Computer Engineering, Virginia Commonwealth University, 
Richmond, VA 23284, USA \\
$^3$Department of Mechanical and Nuclear Engineering, Virginia Commonwealth University, 
Richmond, VA 23284, USA}
\date{\today}

\date{\today}

\begin{abstract}
A longstanding goal of spintronics is to inject, coherently transport, and 
detect spins 
in a semiconductor nanowire where a {\it single} quantized subband is occupied at room temperature. Here,
we report achieving this goal
in 50-nm diameter 
InSb nanowires by demonstrating both the spin-valve and 
the Hanle effect. 
The spin relaxation time in the nanowires was found to have increased by an order of magnitude 
over what has been
reported in bulk and quantum wells due to the suppression of 
D'yakonov-Perel' spin relaxation as a result of single subband occupancy. 
These experiments raise hopes for the realization of a room-temperature
 Datta-Das spin transistor.
\end{abstract}

\pacs{72.25.Dc, 72.25.Hg, 72.25.Rb}
\keywords{spin transport, nanowires, Hanle effect, spin valves}
\maketitle


Two phenomena that unambiguously demonstrate injection, coherent transport and 
detection of spins in a solid are the spin-valve effect and the Hanle effect.
In the former effect, the magnetoresistance of a
trilayered structure, consisting of 
a paramagnet interposed
between two ferromagnets that inject and detect spins, exhibits either 
a peak or a trough between the coercive fields 
of the two ferromagnets \cite{julliere}. In the latter effect,
the conductance of the structure exhibits periodic 
oscillations in a magnetic field that is non-collinear with 
the magnetizations of the two 
ferromagnets. Both  effects have been observed simultaneously at 
low temperature \cite{lou} and room temperature \cite{kum},
but in systems where multiple
quantized electron states (subbands) are occupied by electrons. Here, we report observing them in a solid 
where a single subband is occupied. Single 
subband occupancy is important for two reasons; it allows complete suppression of the dominant
spin relaxation mechanism (D'yakonov-Perel') in most semiconductors \cite{dyakonov, holleitner, pramanik1}
and it also eliminates wavevector dependence of the 
spin precession thereby nullifying the deleterious effects of ensemble averaging at
a finite temperature. As a consequence, a Datta-Das spin transistor \cite{datta}, implemented with a single 
subband occupied quantum wire channel, can ideally produce infinite conductance on/off
ratio at room temperature. No other structure can do this.

Parallel  arrays of trilayered nanowires, each of $\sim$50 nm diameter and 
consisting of 
Co, InSb and Ni layers, were fabricated by sequentially electrodepositing Co, InSb 
and Ni in 50-nm
diameter pores of nanoporous anodic alumina films. Details of 
fabrication are given in the supplementary material. Each nanowire is a ``spin-valve''; 
one ferromagnetic contact injects spin-polarized 
electrons via tunneling through the Schottky barrier formed at the ferromagnet/semiconductor interface, 
and the other ferromagnetic contact transmits the electrons to varying degrees depending on their
spin polarization. Thus, the injecting contact acts as a spin polarizer and the detecting contact acts
as a spin analyzer. In each nanowire, $\sim$96\% of the 
electrons occupy the lowest subband at room temperature (see supplementary material) so that there is essentially
single subband occupancy. Approximately
10$^{8}$ parallel wires are electrically 
contacted from the top and bottom, forming an assembly of 10$^{8}$ parallel resistors. A schematic of the structure 
is shown in the inset of Fig. 1 which shows a transmission electron micrograph of a single
trilayered nanowire that formed within the pores. Energy dispersive analysis of x-ray 
(EDAX) spectra of the samples are shown in the supplementary material. Cross-section scanning electron micrographs 
of similar structures can be found in ref. [\onlinecite{saumil}], showing the wires' geometry.

\begin{figure}
\includegraphics[width=3in]{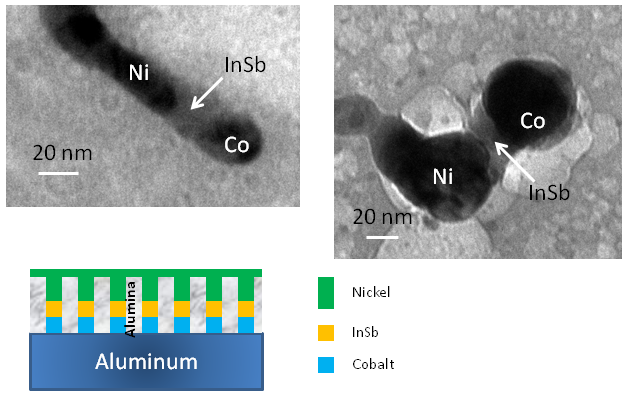}
\caption{\label{sample} Transmission electron micrograph of two trilayered nanowires
that had formed within the pores. The nanowires were first released from their alumina host by
dissolving alumina in chromic/phosphoric acid at 70$^{\circ}$C. Next, they
were  captured on carbon 
grids by soaking the grids in the acid solution. The nanowires 
dispersed on the grids were then imaged. The inset
shows a schematic of a slightly overfilled sample's cross-section.}
\end{figure}

Longitudinal magnetoresistance (magnetic field directed 
parallel to wire axes) plots were obtained for three different samples at room 
temperature and are shown in 
Fig. 2. Each sample
exhibits a clearly discernible trough in the magnetoresistance during both 
forward and reverse 
scans of the field. In samples 1 and 
 3, the troughs occur at magnetic fields that are consistent 
 with coercivity values  reported in the 
literature for cobalt and nickel nanowires produced in anodic alumina 
pores \cite{zeng,zheng}. This confirms that the troughs are indeed
due to the spin-valve effect \cite{alphenaar}. In sample 2, however, the troughs occur at 
very low fields, which can only happen if the coercivity of the nickel layer 
has dropped $\sim$30-fold to $\sim$50 Oe. This low 
coercivity may be due to excessive overfilling of the anodic alumina 
pores during nickel 
electrodeposition,
resulting in the formation of a thick nickel layer on the surface that behaves like bulk
instead of a nanowire. 
The coercivity of 
bulk nickel is only $\sim$20 Oe \cite{geiss}. Scanning
electron microscopy confirmed that indeed this had happened 
in some samples (see supplementary material).

Observation of the spin valve 
effect establishes successful injection, coherent transport and
detection of spins at room temperature. 
Since cobalt and nickel both have negative spin polarizations 
\cite{tsymbal},
one would have expected the spin-valve effect to produce a magnetoresistance
 {\it peak} instead of a {\it trough}. In reality, what we are observing here is the 
 {\it inverse} spin valve effect that has been seen before and explained 
 \cite{tsymbal1,pramanik}. The sign inversion of the peak (which makes it
 a trough)
is believed to be caused by electrons resonantly tunneling through one or more localized defect sites in
the InSb spacer layer whose energies match the Fermi energy of the ferromagnetic electrodes. This effectively
inverts the spin polarization of the ferromagnet closer to the defect site \cite{tsymbal1}. Such behavior has been 
observed previously in ferromagnetic/paramagnetic nanojunctions of cross section smaller than 0.01 $\mu m^2$ grown 
by electrodeposition, as is the case here \cite{tsymbal1}.

\begin{figure}
\includegraphics[width=2.7in]{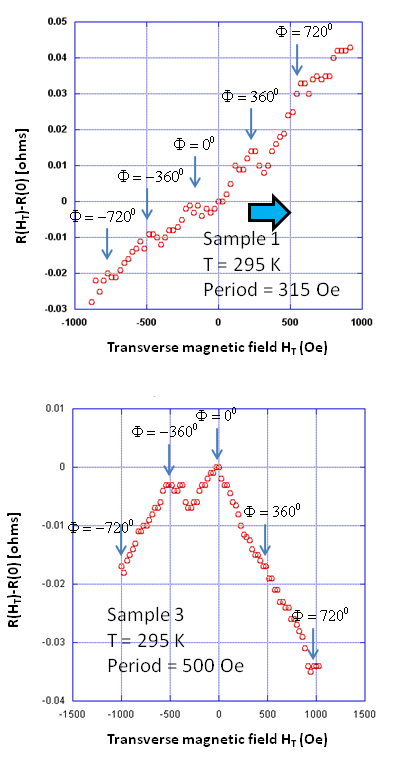}
\caption{\label{spin-valve} {\bf The inverse spin valve effect.} Room temperature (295 K) longitudinal 
magnetoresistance plots of three samples
showing spin valve troughs.  The magnetic field is applied along the axes of the 
nanowires. The trough positions are indicated with vertical arrows.
The horizontal block arrows show the directions in which the field is scanned.
The troughs are not symmetric about the resistance axis 
because of the inevitable asymmetric shapes of the magnets, which make the coercivities 
of both cobalt and nickel contacts depend on the field direction.}
\end{figure}

Each plot in Fig. 2 shows a monotonically increasing background magnetoresistance.
This is expected 
in narrow gap semiconductors like InSb owing to the strong non-parabolicity of the conduction
band \cite{aurora}. In the case of sample 1, however, there was also a thermal drift in the 
resistance due to sample heating by the current, which 
contributes part of the observed background magnetoresistance. No such drifts were 
observed in samples 2 and 3.

One can relate the magnitude of the trough $\Delta R$ to the spin relaxation length in
the InSb spacer layer by invoking the modified Julliere formula \cite{vardeny} which is
\begin{equation}
{{\Delta R}\over{R(0)}} = {{2 P_2 P_2 e^{-L/L_s}}\over{1 - P_1 P_2 e^{-L/L_s}}} ,
\label{jull1}
\end{equation}
where $R(0)$ is the resistance of the structure at zero magnetic field, $P_1$ and $P_2$
are the spin polarizations of the two magnetic contacts, $L$ is the separation
between the contacts and 
$L_s$ is the spin relaxation length.

In our nanowires, 96\% of the electrons reside in the 
lowest subband and hence we have almost a true 
one-dimensional system (see Sections S6 and S7 of the supplementary material for theoretical and 
experimental substantiation of single-subband occupancy). Therefore,
we can write $L = v_d \langle \tau_t \rangle $ and $L_s = v_d \langle \tau_s \rangle $, 
where $v_d$ is the drift velocity of electrons, 
$\langle \tau_t \rangle $ is the ensemble averaged 
transit time and $\langle \tau_s \rangle$ is the ensemble averaged spin relaxation time. 
This reduces Equation
(\ref{jull1}) to 
\begin{equation}
{{\Delta R}\over{R(0)}} = {{2 P_2 P_2 e^{-\langle 
\tau_t \rangle/
\langle \tau_s \rangle}}\over{1 - P_1 P_2 e^{-\langle 
\tau_t \rangle/
\langle \tau_s \rangle}}} ,
\label{jull2}
\end{equation}

The spin polarizations in bulk nickel and cobalt at 0 K have been reported as 0.33 and 0.42, 
respectively 
\cite{tsymbal}, but at room temperature, the value reported
in 50 nm thin films of cobalt is only 0.1 \cite{jedema2}. There is one report that the 
value increased to 0.16 when the 
resistance of the junction between cobalt and a non-magnetic spacer increased
\cite{tinkham}, but this is not relevant to our work. 
Furthermore, interface effects always reduce spin polarization \cite{dowben},
and we have a very large interface-to-volume ratio in our cobalt nanocontacts. 
Therefore, it is reasonable to expect that
  the spin polarization in
 our cobalt nanocontacts is no more than 0.1. There is no equivalent study for 
nickel, but we will assume a similar 4-fold reduction in the spin polarization and take it
to be no more than 0.075 in the nickel nanocontacts. Based on these spin polarization values 
and the measured 
resistance ratio  ${{\Delta R}\over{R(0)}}$, we infer that the ratio $\langle \tau_t \rangle/
\langle \tau_s \rangle $ is no more than 3.4 in sample 1,
2.8 in sample 2 and 3.6 in sample 3 at room temperature.

The transverse magnetoresistances of the samples are then measured in a magnetic field  
perpendicular to the axes of the wires. As long as this 
perpendicular field does not exceed the coercivities of the cobalt and nickel 
nanocontacts,
the contacts will continue to inject and extract spins polarized parallel to the wire axis.
These spins will precess about the perpendicular field while traversing the InSb layer. If the angle by which the majority 
spins precess in transiting the InSb layer is an odd multiple of 180$^{\circ}$, they
will be transmitted since the spin polarizations of the two contacts are effectively opposite
in sign (spin valve `trough'). On the other hand, if the angle is an even multiple of 180$^{\circ}$,
the majority spins will be blocked.
Hence, the resistance of the sample will oscillate 
periodically as the transverse magnetic field is scanned 
since the angle of precession is proportional to the transverse field. 
This is the well-known 
Hanle effect \cite{jedema2, takamura, appelbaum, fukuma}. The period of the oscillation in magnetic flux 
density should be $B_{period} = h/(|g| \mu_B \langle \tau_t \rangle)$, where $\mu_B$ is the 
Bohr magneton and $g$ is the Land\'e g-factor of the spacer material (InSb).

\begin{figure}
\includegraphics[width=3in]{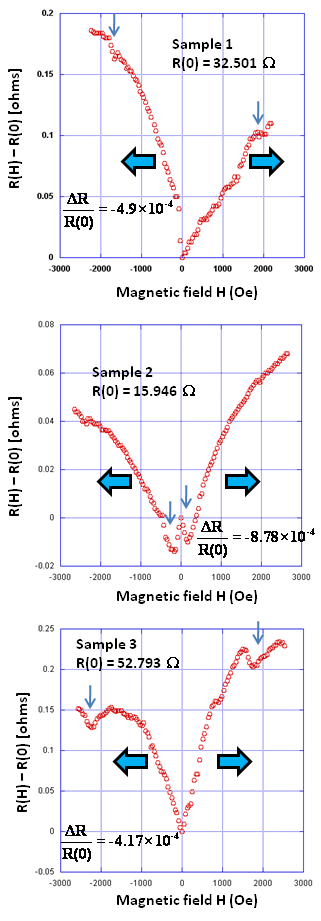}
\caption{\label{hanle} {\bf Hanle oscillations}. Room temperature (295 K) transverse 
magnetoresistance plots of two samples showing
 oscillations due to the Hanle effect.  
The resistance maxima are indicated by arrows and the angle of spin precession 
is the quantity $\Phi$. In sample 1, there is a zero-offset in $\Phi$ and the maximum
closest to zero field is chosen for $\Phi = 0$. The block arrow shows the direction of field 
scan.}
\end{figure}

In Fig. 3, we show the transverse magnetoresistance plots of samples 1 and 3 
 at room temperature. The transverse magnetic field strength is kept well below the coercivities of 
the cobalt and nickel nanocontacts (estimated from the longitudinal 
magnetoresistance plots in Fig. 2) in order 
to avoid flipping their 
magnetizations. Hence, their magnetizations remain parallel to the wire axis 
and
they continue to inject and detect majority spins with polarization 
along the 
wire axis which is perpendicular to the applied field. 

In sample 1, there is a distinct
periodic oscillation in the resistance or conductance (equally spaced minima and equally spaced maxima)
with a period of $\sim$315 Oe. 
This is the signature of the Hanle effect.
In the case of sample 3,
the oscillation is more muted, but there is still a hint of oscillation with a
 period of $\sim$ 500 Oe. 
Sample 2 did not show any oscillation since the 
coercivity of its nickel contact is too low for the observation of the 
Hanle effect. The oscillations are clearly non-sinusoidal because different nanowires produce 
slightly different periods, and ensemble averaging over the periods distorts the shape of the
oscillation, making it appear non-sinusoidal.

There are two sources that introduce a spread in the 
transit time $\tau_t$ and hence a spread in the period of the Hanle oscillation. The first is
inhomogeneous broadening due to a spread in the spacer layer width $L$ among the different nanowires, and the 
second is homogeneous broadening due to a spread in the electron velocity even within the same nanowire.
As long as the distribution in the spacer layer width and carrier velocity (and hence the distribution in
the period) is somewhat peaked, we will be able to observe a periodic behavior (with the period corresponding to
the peak in the distribution), but the oscillation will be non-sinusoidal.
In other words, successive minima and maxima will be equally spaced in the magnetic field,
but the minima may not occur exactly midway between two successive 
maxima and vice versa. This is precisely what we see.
The homogeneous broadening is fortunately small and the spread in velocity is much 
less than the thermal spread since electrons resonantly tunnel through one or more defect sites (which is 
why we observe the inverse spin valve effect) and resonant tunneling acts as a velocity filter. 
Only electrons whose energies are very close to the resonant level can tunnel through, which
reduces the spread in kinetic energy and electron velocity. Therefore, most of the spread in the oscillation period accrues 
from the inhomogeneous broadening.
The variation in the spacer layer width is 25\% - 35\% (see supplementary material), which should result in a 
similar variation in the period. This variation distorts the oscillation,
making it non-sinusoidal, but still discernible.

Notice that the Hanle oscillations are not symmetric about zero transverse magnetic field. This usually happens when the transverse
magnetic field is slightly misaligned and not exactly perpendicular to the axis of the nanowire \cite{appelbaum_arxiv}. 
 In our samples, we can never align the magnetic field 
exactly perpendicular to every wire's axis since the wires themselves are not exactly mutually parallel.
Hence, we will always observe some symmetry-breaking.
Note also that in sample 1, 
the zero-field 
resistance is not a maximum and that there is a shift. This happens when the magnetizations of the 
two contacts are not exactly 
collinear.


Both samples show a background transverse 
magnetoresistance.
In the case of sample 1, the resistance increases with time due to 
thermal drift and hence increases as the magnetic field is scanned 
slowly from negative to positive values. In the case of sample 3, which had no thermal drift,
the magnetoresistance is negative and it is most likely caused
by increasing depopulation of the higher subbands as the transverse field is 
increased. 
The transverse field 
increases the energy separation between successive subbands and therefore increases 
the population of the lowest subband at the cost of higher subband populations. The lowest 
subband usually has the highest mobility since its wavefunction is peaked at the center of the 
nanowire. Thus, one expects a negative transverse magnetoresistance in a nanowire. 
In our samples,
96\% of the electrons are already in the lowest subband and only 4\% are in higher subbands,
which should make this effect almost negligible.
Since the magnetoresistance is only 0.05\%, it is quite plausible that this tiny decrease
in the resistance in the transverse field is a consequence of higher subband depopulation.

From the observed average period $B_{period}$
in the magnetic flux density, one can calculate the ensemble averaged transit time 
of electrons through the InSb spacer layer according to the relation
\begin{equation}
\langle \tau_t \rangle = {{h}\over{|g| \mu_B B_{period}}} .
\end{equation}

The g-factors of materials are known to change in nanostructures from the bulk value,
but in InSb quantum dots, the g-factor has been reported to be $-$52 in the lowest subband \cite{nilsson}, 
which is close 
to the bulk value of $-$51. Based on this, we calculate $\langle \tau_t \rangle $ = 44 ps 
in sample 1
and 28 ps in sample 3. Therefore, the ensemble averaged spin 
relaxation time $\langle \tau_s \rangle $ in samples 1 and 3 are 
at least 13 ps and 8 ps, respectively. Note that these are true {\it transport}
spin lifetimes, since they are measured in a transport experiment. Previous measurements 
of spin lifetimes have been carried out with optical techniques and not 
transport methods \cite{litvinenko,murzyn,bhowmick}, which may not yield the true transport 
lifetime.

The room temperature spin relaxation time in {\it epilayers} of InSb has been reported as 
low as 2.5 ps \cite{litvinenko}
and as high as 300 ps \cite{murzyn}, but in quantum wells, it drops to 1.2 - 4.2 ps \cite{bhowmick}.
It has been suggested theoretically that intrinsic inversion symmetry breaking
at the interfaces of a quantum well can decrease spin relaxation time by over 
an order of magnitude
\cite{olesberg}. The calculated spin relaxation time in InAs/GaSb quantum wells
 is only 0.9 ps
\cite{olesberg}. In InSb nanowires of 50 nm diameter, we will expect the spin 
relaxation time to be even shorter than in quantum wells because of 
the larger interface-to-volume ratio, and hence shorter than
1 ps at room temperature. 
Therefore, the measured relaxation times of 13 and 8 ps are roughly an order 
of magnitude {\it longer} than what is expected.

The electron drift mobility in the InSb nanowires was ascertained from TEM observation of the spacer layer width,
measured sample resistance and 
transit time extracted from the measured Hanle period (see Section S3 of supplementary material). 
It was found to have degraded by four orders of magnitude
from the value reported in bulk or 
quantum wells,
presumably owing to severe interface roughness scattering brought about by the nanowires having a large 
interface-to-volume ratio. The mobility degradation should increase the Elliott-Yafet spin relaxation 
rate by four orders 
of magnitude compared to bulk or quantum wells since this rate is inversely proportional to mobility \cite{elliott}. 
Yet, the spin relaxation rate had
{\it decreased} by nearly
one order of magnitude. This can only happen if: (1) Elliott-Yafet is not the dominant spin relaxation 
mechanism in bulk and quantum wells of InSb, but D'yakonov-Perel'  is  (this was theoretically
predicted in ref. [\onlinecite{kim}]), and (2) the D'yakonov-Perel' mechanism has been 
suppressed or eliminated in the nanowires because of strictly
one dimensional confinement of carrier motion resulting from single subband occupancy \cite{holleitner, pramanik1}.
In Section S6 of the supplementary material, we present calculations to show that we indeed have single subband occupancy 
in our nanowires, and in section S7 we show that the room temperature conductance can be used to estimate the 
number of occupied subbands and it is, again, indeed O(1).

In conclusion, we have demonstrated both the spin valve and the Hanle effect at room temperature 
in trilayered nanowires of Co-InSb-Ni, thereby demonstrating spin injection,
coherent spin transport and spin detection in a 
semiconductor at room temperature. The material InSb has a strong Rashba coefficient 
\cite{rashba},
which makes it an ideal candidate for the Datta-Das spin field effect transistor \cite{datta}. 
This transistor requires a strictly one-dimensional channel with a single subband occupied for the strongest 
effect \cite{book}; hence, the demonstration of coherent spin transport in these single-subband-occupied 
InSb nanowires -- where the D'yakonov-Perel' spin relaxation has been suppressed -- 
raises hopes for a room temperature device with significant conductance modulation.

This work is supported by the US National Science Foundation under grant 
CCF 0726373. The authors are indebted to Dr. Dmitry Pestov for help with EDAX.


\end{document}